\documentclass[%
reprint,
superscriptaddress,
showpacs,preprintnumbers,
amsmath,amssymb,
prb,
]{revtex4-1}
\usepackage{graphicx}
\usepackage{dcolumn}
\usepackage{xfrac}
\usepackage{bm}
\usepackage{color}
\usepackage[export]{adjustbox}
\usepackage{xr-hyper}
\usepackage{hyperref}
\usepackage{tabularx}

\newcommand{\FC}{FeCl$_3$}
\newcommand{\FB}{FeBr$_3$}
\newcommand{\FCB}{FeCl$_{3-x}$Br$_x$}

\newcommand{\TN}{$T_\textrm{N}$}
\newcommand{\TW}{$\Theta_\textrm{W}$}
\begin{document}

\title{Extreme sensitivity of the magnetic ground-state to halide composition in \FCB}

\author{Andrew~Cole}
\affiliation{Department of Physics, Boston College, Chestnut Hill, MA 02467, USA}

\author{Alenna~Streeter}
\affiliation{Department of Physics, Boston College, Chestnut Hill, MA 02467, USA}

\author{Adolfo~O.~Fumega}
\affiliation{Department of Applied Physics, Aalto University, Espoo, Finland}

\author{Xiaohan~Yao}
\affiliation{Department of Physics, Boston College, Chestnut Hill, MA 02467, USA}

\author{Zhi-Cheng~Wang}
\affiliation{Department of Physics, Boston College, Chestnut Hill, MA 02467, USA}

\author{Erxi~Feng}
\affiliation{Neutron Scattering Division, Oak Ridge National Laboratory, Oak Ridge, Tennessee 37831, USA}

\author{Huibo~Cao}
\affiliation{Neutron Scattering Division, Oak Ridge National Laboratory, Oak Ridge, Tennessee 37831, USA}

\author{Jose~L.~Lado}
\affiliation{Department of Applied Physics, Aalto University, Espoo, Finland}

\author{Stephen~E.~Nagler}
\affiliation{Neutron Scattering Division, Oak Ridge National Laboratory, Oak Ridge, Tennessee 37831, USA}

\author{Fazel~Tafti}
\affiliation{Department of Physics, Boston College, Chestnut Hill, MA 02467, USA}


\begin{abstract}
Mixed halide chemistry has recently been utilized to tune the intrinsic magnetic properties of transition-metal halides -- one of the largest families of magnetic van der Waals materials.
Prior studies have shown that the strength of exchange interactions, hence the critical temperature, can be tuned smoothly with halide composition for a given ground-state.
Here we show that the ground-state itself can be altered by a small change of halide composition leading to a quantum phase transition in \FCB.
Specifically, we find a three-fold jump in the N\'{e}el temperature and a sign change in the Weiss temperature at $x= 0.08$ corresponding to only $3\%$ bromine doping.
Using neutron scattering, we reveal a change of the ground-state from spiral order in \FC\ to A-type antiferromagnetic order in \FB.
Using first-principles calculations, we show that a delicate balance between nearest and next-nearest neighbor interactions is responsible for such a transition.
These results support the proximity of \FC\ to a spiral spin liquid state, in which competing interactions and nearly degenerate magnetic $k$-vectors may cause large changes in response to small perturbations.
\end{abstract}

\maketitle


\section{\label{sec:introduction}Introduction}
Magnetic frustration provides a fascinating playground for the realization of exotic quantum states\cite{savary_quantum_2016}.
A curious example of frustrated magnet is the spiral spin-liquid (SSL) phase produced by competing interactions in a bipartite lattice such as diamond (3D) and honeycomb (2D) structures.
The SSL is characterized by degenerate spin spirals with $\mathbf{k}$-vectors lying on a surface in momentum space~\cite{yao_generic_2021}.
Weak thermal fluctuations can entropically lift this degeneracy and establish order by disorder~\cite{villain_order_1980}.
Spinel materials such as MnSc$_2$S$_4$ and CoAl$_2$O$_4$ are candidates of SSL in the 3D diamond lattice described by a frustrated $J_1$-$J_2$ Heisenberg model~\cite{fritsch_spin_2004,macdougall_kinetically_2011,gao_spiral_2017,bergman_order-by-disorder_2007,niggemann_classical_2019}.
In 2D, however, the experimental realization of such effects in the honeycomb lattice~\cite{mulder_spiral_2010,zhang_exotic_2013}, prevalent in van der Waals (VdW) materials, has remained elusive.

Recent neutron scattering experiments on the VdW magnet \FC\ with a honeycomb lattice have shown a ring of degenerate $\mathbf{k}$-vectors just above \TN$=8.5$~K, indicating a 2D SSL phase~\cite{gao_spiral_2022}.
Below \TN, a spiral order with $\mathbf{k}=(\frac{4}{15},\frac{1}{15},\frac{3}{2})$ is established~\cite{cable_neutron-diffraction_1962,jones_lowtemperature_1969} indicating the entropic selection of this propagation vector by spin fluctuations, i.e. order by disorder.

In this letter, we reveal the extreme sensitivity of the ground-state of \FC\ to tiny amounts of disorder by characterizing a series of \FCB\ crystals (Fig.~\ref{fig:CIF}a--c).
We find a three-fold jump in the N\'{e}el temperature (\TN) and a sign change in the Weiss temperature (\TW) between $x=0$ and $0.08$ corresponding to only $3\%$ bromine doping.
Such dramatic enhancement of the order due to a tiny amount of disorder suggests that \FC\ is a frustrated magnet on the verge of a quantum phase transition (QPT) between the SSL phase and a different ordered state (Fig.~\ref{fig:CIF}d).
For $x\ge 0.08$, we identify the ordered state to be A-type AFM which is ferromagnetic (FM) within the layers and AFM between them.
This is similar to the ground-state of \FB\ and different from the spiral order in \FC.

Our experimental results are corroborated by first-principles calculations on a 2D $J_1$-$J_2$ Heisenberg model that predict a QPT between the SSL and FM states in \FCB\ as observed experimentally.
We discuss the importance of including $p$-orbital correlations, in addition to $d$-orbital correlations, for obtaining the correct magnetic ground-state in density functional theory (DFT).
To our knowledge, the impact of $p$-orbital correlations in DFT calculations for VdW materials has not be discussed in the literature before. 
Details of crystal growth, neutron scattering, scanning electron microscopy (SEM), magnetization measurements, and DFT calculations are explained in the Supplementary Information (SI).

\section{\label{sec:results}Results and Discussion}
\begin{figure}
  \includegraphics[width=0.46\textwidth]{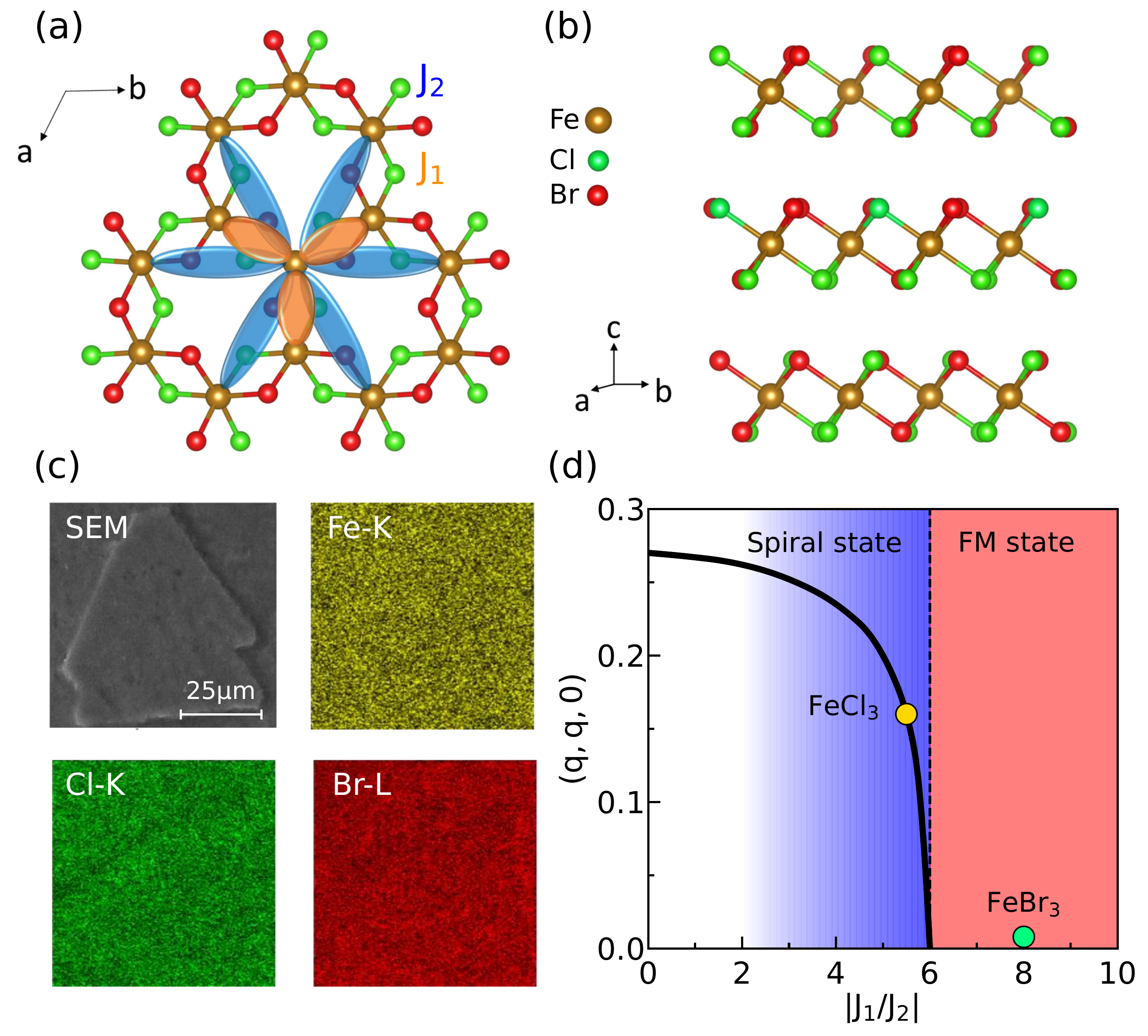}
  \caption{\label{fig:CIF}
  (a) Schematic illustration of the honeycomb $ab$-planes in the mixed halide system FeCl$_{1.5}$Br$_{1.5}$.
  The $J_1$ and $J_2$ exchange paths are highlighted in orange and blue colors, respectively.
  (b) Layered (VdW) structure of FeCl$_{1.5}$Br$_{1.5}$ viewed from the [210] direction.
  (c) SEM image and EDX color maps confirming the homogeneous distribution of halides in an FeCl$_{1.33}$Br$_{1.67}$ crystal.
  (d) Phase diagram of the spiral and FM states with \FC\ and \FB\ across the QPT.
  Here, the $(q,q,0)$ $y$-label is the degenerate magnetic wave vector for the SSL state, not the final spiral ground-state~\cite{gao_spiral_2022}.  
  \FB\ is not in the SSL phase, and its order is FM in 2D. 
  }
\end{figure}
Among different classes of VdW magnets, transition-metal halides offer a special opportunity in that their intrinsic properties can be tuned by mixing the halide species (Cl, Br, and I)~\cite{abramchuk_controlling_2018}.
With increasing halide size, the orbital overlaps and ligand spin-orbit coupling (SOC) are enhanced, which in turn tune all magneto-optical properties as reported previously in CrCl$_{3-x}$Br$_x$, CrBr$_{3-y}$I$_y$, and CrCl$_{3-x-y}$Br$_x$I$_y$ alloys~\cite{abramchuk_controlling_2018,tartaglia_accessing_2020}.

Following the recent report of a SSL ground-state and order by disorder in \FC,~\cite{gao_spiral_2022} we grew the heteroanionic crystals of \FCB\ with the goal of tuning the SSL ground-state.
Since both \FC\ and \FB\ crystallize in the rhombohedral space group $R\bar{3}m$ with a layered honeycomb structure, a solid solution of \FCB\ crystals can also be grown with the same structure (Figs.~\ref{fig:CIF}a,b). 
The composition and uniform distribution of elements were confirmed using energy dispersive x-ray spectroscopy (EDX).
For example, Fig.~\ref{fig:CIF}c shows the distribution of Fe $K$-edge (yellow), Cl $K$-edge (green), and Br $L$-edge (red) absorption intensities in FeCl$_{1.33}$Br$_{1.67}$.

The super-exchange interactions between Fe$^{3+}$ ions are mediated by Cl and Br ligands within the honeycomb layers of each \FCB\ sample.
The nearest neighbor (NN) and next-nearest neighbor (NNN) exchange paths are denoted by $J_1$ and $J_2$ in Fig.~\ref{fig:CIF}a, respectively.
Although $J_2$ is weaker than $J_1$, there are six NNNs and three NNs, leading to a close competition between the $J_1$ and $J_2$ terms in the model Hamiltonian
\begin{equation}
\label{eq:J1J2}
H = 
J_1 \sum_{\langle ij \rangle} \vec S_i \cdot \vec S_j
+
J_2 \sum_{\langle\langle ij \rangle\rangle} \vec S_i \cdot \vec S_j
\end{equation}
where $S=5/2$ in the high-spin state of Fe$^{3+}$ ($^{6}S_{5/2}$).
The competition between different magnetic exchange paths (direct and super-exchange) leads to effective $J_1<0$ (FM) and  $J_2>0$ (AFM) for the 2D spin model of Eq.~\ref{eq:J1J2}. 
The ratio between $J_1$ and $J_2$, $\left|J_1 / J_2\right|$, determines the magnetic ground-state of the honeycomb lattice as well as the degenerate $(q,q,0)$ wave vector of the SSL phase above \TN\ which is shown schematically in Fig.~\ref{fig:CIF}d. 
It is known from prior neutron diffraction experiments~\cite{cable_neutron-diffraction_1962} that spin spiral is the ground-state of \FC. 
We will show here that the Br substitution modifies $|J_1 / J_2|$ ratio and drives the system to an A-type AFM state (which is FM in 2D) where \FB\ is located in Fig.~\ref{fig:CIF}d.

\subsection{\label{subsec:Magnetization}Magnetization Measurements} 
\begin{figure*}
  \includegraphics[width=\textwidth]{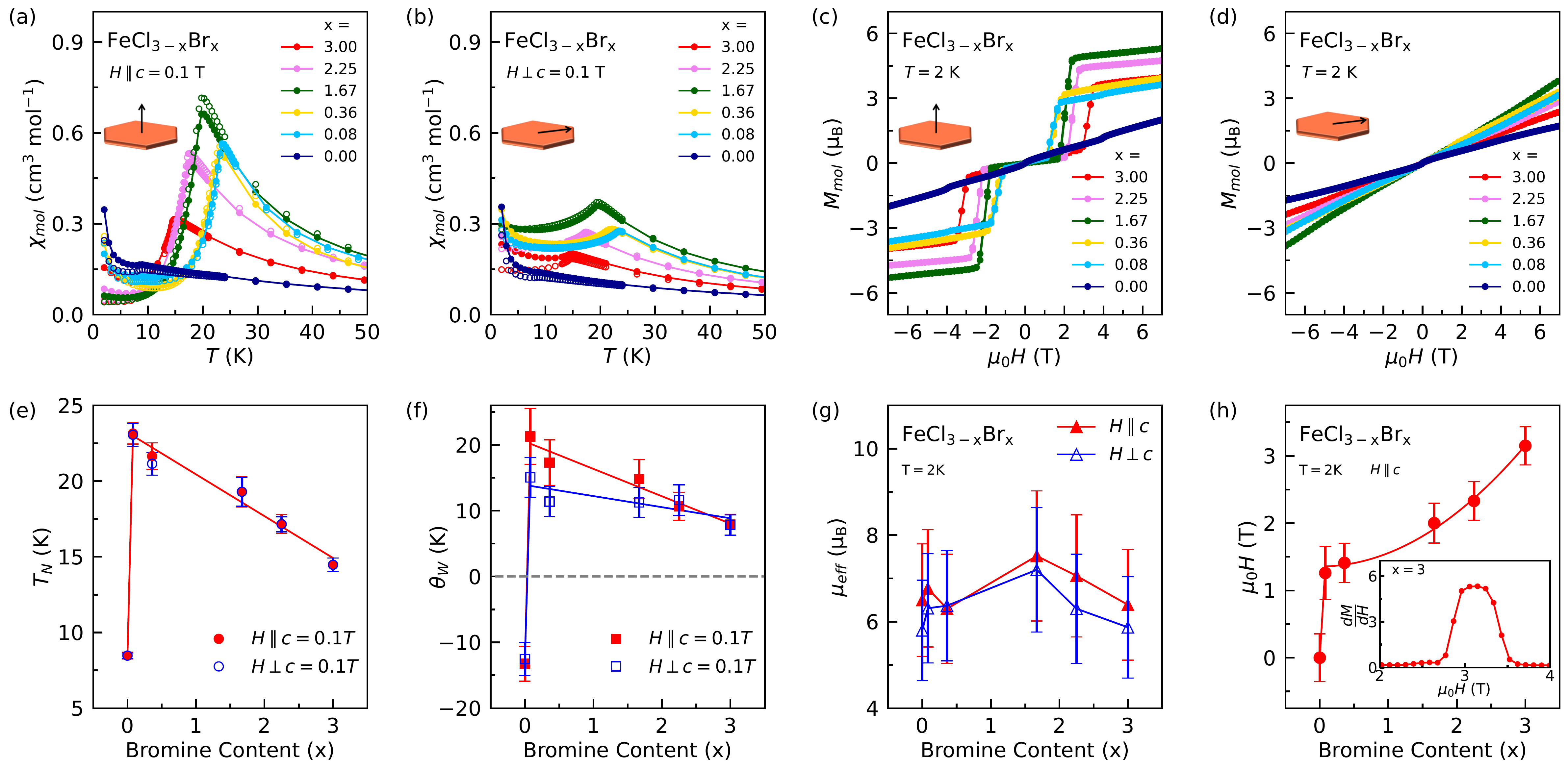}
  \caption{\label{fig:MAG}
  (a) Magnetic susceptibility as a function of temperature measured under ZFC (open circles) and FC (full circles) conditions with $H\| c$.
  (b) Same as in (a) but with $H\perp c$.
  (c) Magnetization curves with $H \| c$ showing metamagnetic (MM) transitions in \FCB\ samples with $x\ge 0.08$.
  (d) The MM transitions are absent when $H \perp c$.
  (e) \TN\ as a function of Br content ($x$) showing an initial jump followed by a linear decrease.
  (f) \TW\ as a function of $x$ showing an initial sign change followed by a linear decrease.
  (g) $\mu_\text{eff}$ estimated from the Curie-Weiss analysis (Fig.~S1).
  Error bars in panels (e,f,g) are mainly due to the uncertainty in evaluating the mass of thin VdW crystals.
  (h) The critical field of MM transitions as a function of $x$ showing an initial jump followed by a smooth increase.
  Error bars reflect the width of the transition (inset).
  }
\end{figure*}
The experimental evidence of competing FM and AFM interactions appear in the magnetic susceptibility ($\chi$) and magnetization ($M$) data.
The raw data for all samples with $H\|c$ and $H\perp c$ are shown in Figs.~\ref{fig:MAG}a-d, and the analyzed results are presented in Figs.~\ref{fig:MAG}e-h.
The empty and full circles in Figs.~\ref{fig:MAG}a,b correspond to zero-field-cooled (ZFC) and field-cooled (FC) measurements, respectively.
In Fig.~\ref{fig:MAG}a, the peaks in $\chi(T)$ curves with $H\| c$ and the minimal difference between the ZFC and FC curves are characteristics of AFM transitions.
Unlike $H\| c$, the $H\perp c$ curves in Fig.~\ref{fig:MAG}b do not go to zero when $T\to0$, indicating a finite FM component.
A combination of FM and AFM correlations also exists in CrCl$_3$, which undergoes an A-type AFM order (FM within the layers and AFM between them)~\cite{cable_neutron_1961}.
In CrCl$_3$, only the $\mathrm{t_{2g}}$ manifold of Cr$^{3+}$ is at half-filling, whereas both $\mathrm{t_{2g}}$ and $\mathrm{e_{g}}$ levels are at half-filling in \FC, maximizing the competition between FM and AFM correlations according to Goodenough-Kanamori rules~\cite{goodenough_interpretation_1958,kanamori_superexchange_1959}.

We determined \TN\ from $d\chi/dT$ curves (Supplementary Fig.~S1), and plotted it as a function of bromine content ($x$ in \FCB) for both field directions in Fig.~\ref{fig:MAG}e.
Using a Curie-Weiss (CW) analysis (Supplementary Fig.~S1), we extracted \TW, which is a rough measure of the magnetic correlations, and plotted it as a function of $x$ in Fig.~\ref{fig:MAG}f.
The central observation in Figs.~\ref{fig:MAG}e,f is a jump in both \TN\ and \TW\ when a tiny amount of Br is added to FeCl$_3$, i.e. at $x=0.08$ in \FCB\ corresponding to only $3\%$ Br doping.
The effect is dramatic with \TN\ showing a three-fold jump and \TW\ changing sign, indicating a change of magnetic ground-state at $x=0.08$.
From the CW analysis, we found the effective magnetic moment of all \FCB\ samples to be close to 5.9~$\mu_B$ within experimental errors (Fig.~\ref{fig:MAG}g) as expected for Fe$^{3+}$ in the high-spin state.

After the initial three-fold jump of \TN\ from $8.5(2)$~K at $x=0$ to $23.1(1)$~K at $x=0.08$, it is suppressed linearly to $14.5(5)$~K at $x=3$ (Fig.~\ref{fig:MAG}e).
This behavior is the same for both field directions.
Similarly, after the initial jump of \TW\ from $-13$~K at $x=0$ to $+21$~K at $x=0.08$, it is suppressed linearly to $8$~K at $x=3$ for both field directions (Fig.~\ref{fig:MAG}f).
Two conclusions can be drawn from these observations.
(i) The initial jump in \TN\ and sign change in \TW\ suggest an abrupt change of the magnetic ground-state of \FC\ by a tiny amount in Br doping, indicating its proximity to a QPT.
(ii) The linear decrease of both \TN\ and \TW\ from $x=0.08$ to $3$ suggests that the ground-states of all \FCB\ samples with $x\ge 0.08$ are similar to the ground-state of \FB\ and different from that of \FC.

Another evidence of the change of magnetic ground-state at $x=0.08$ comes from the field dependence of magnetization. 
The $M(H)$ curves of \FC\ in Figs.~\ref{fig:MAG}c,d are nearly linear in both field directions, consistent with the spiral AFM order reported in prior neutron diffraction studies~\cite{cable_neutron-diffraction_1962}.
The subtle kinks in the $M(H \perp c)$ curves at $H=0$ and $4$~T are due to the alignment of spiral domains with the field.
Unlike \FC, the \FCB\ samples with $x\ge 0.08$ exhibit field-induced metamagnetic (MM) transitions when $H\perp c$ and linear $M(H)$ when $H\|c$ (Figs.~\ref{fig:MAG}c,d).
Tracing the critical field ($H_c$) of the MM transition as a function of $x$ (Fig.~\ref{fig:MAG}h) reveals an initial jump at $x=0.08$ followed by a smooth increase of $H_c$ for $x\ge 0.08$. 
The $H_c$ values in Fig.~\ref{fig:MAG}h correspond to the peak fields in the $dM/dH$ curves shown in the inset.
Similar to the behavior of \TN\ and \TW\ (Figs.~\ref{fig:MAG}e,f), the initial jump of $H_c$ at $x=0.08$ in Fig.~\ref{fig:MAG}h indicates a change of the magnetic ground-state, and its subsequent smooth increase indicates that the ground-states of \FCB\ samples with $x\ge 0.08$ are similar to that of \FB\ and different from \FC.

\subsection{\label{subsec:Neutron}Neutron Diffraction.}
\begin{figure*}
  \includegraphics[width=\textwidth]{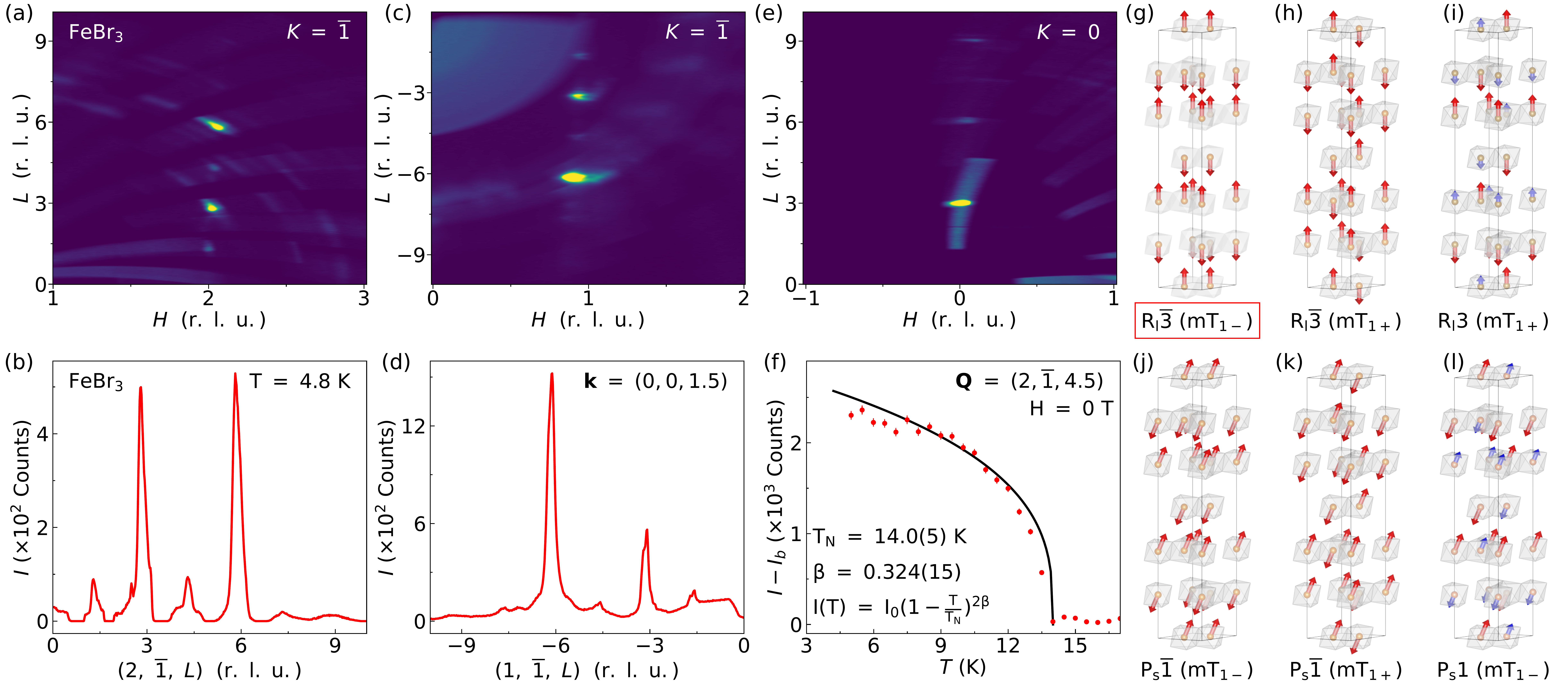}
  \caption{\label{fig:NEUTRON}
  (a) The 2D neutron diffraction scan at 4.8~K and zero-field in the $H$-$L$ plane at $K=\bar{1}$, showing strong nuclear reflections at integer $L$ and weak magnetic reflections at half-integer $L$. 
  (b) A 1D cut through the data in panel (a) showing nuclear (strong) and magnetic (weak) Bragg peaks that identify the magnetic propagation vector $\mathbf{k} = (0, 0, 1.5)$.
  (c) 2D scan in the $(H\bar{1}L)$ plane showing another set of nuclear and magnetic reflections.
  (d) 1D cut through the data in panel (c) showing Bragg peaks along $(1\bar{1}L)$ direction.
  (e) 2D scan in the $(H0L)$ plane showing the absence of magnetic reflections along $(00L)$ direction.
  (f) Temperature dependence of the intensity of the magnetic reflection $\mathbf{Q} = (2, \bar{1}, 4.5)$ with a power-law fit to extract \TN\ and the critical exponent $\beta$. 
  (g,h,i) Candidate magnetic ground-states of \FB\ with $R_l$ symmetry.
  (j,k,l) Candidate ground-states with $P_s$ symmetry.
  The ground-state is determined as A-type AFM shown in panel (g).
  }
\end{figure*}
Single crystal neutron diffraction has been performed on \FC\ previously, and the ground-state was determined as AFM with a spiral modulation parallel to the $[14\bar{5}]$ direction with a wavelength of 15 $(14\bar{5})$ $d$ spacing~\cite{cable_neutron-diffraction_1962}.
We performed single-crystal neutron diffraction on \FB\ at 4.8~K and zero field to probe its magnetic ground-state.
Since all \FCB\ samples with $x\neq 0$ have similar $\chi(T)$ and $M(H)$ behaviors as \FB\ (Fig.~\ref{fig:MAG}), we assume that their magnetic ground-states are similar to that of \FB.

The observed Bragg peaks in Figs.~\ref{fig:NEUTRON}a,b at $\mathbf{Q}=(2,\bar{1},L)$ consist of structural (nuclear) peaks at $L=3$ and $6$ as well as magnetic peaks at $L=\frac{3}{2}$, $3+\frac{3}{2}$, and $6+\frac{3}{2}$, hence the magnetic propagation vector $\mathbf{k}=\left(0,0,1.5 \right)$.
The systematic absence of the nuclear peaks other than $L=3n$ ($n\in$ integer) in Figs.~\ref{fig:NEUTRON}a,b is expected in the space group $R\bar{3}$ (\#148) of \FB.
A second scan along $\mathbf{Q}=(1,\bar{1},L)$ in Figs.~\ref{fig:NEUTRON}c,d confirms the propagation vector $\mathbf{k}=\left(0,0,1.5 \right)$.
The absence of magnetic reflections at $\mathbf{Q}=(0,0,4.5)$ and $(0,0,7.5)$ in Fig.~\ref{fig:NEUTRON}e suggests that the ordered magnetic moments lie along the $c$-axis because neutrons only probe the moment perpendicular to scattering vector ($\mathbf{M}\perp \mathbf{Q}$).
Thus, we identify \FB\ as an out-of-plane Ising system with $M_{z}\neq0$ and $M_{x,y}=0$.

We construct an order parameter plot in Fig.~\ref{fig:NEUTRON}f by tracing the intensity of the $\mathbf{Q}=(2,\bar{1},4.5)$ peak as a function of temperature. 
From a power-law fit, we extract $T_\text{N} = 14.0(5)$~K consistent with $14.5(5)$~K from the magnetization measurements, and $\beta=0.324(15)$ consistent with a 3D Ising system.

A magnetic symmetry analysis based on the ordering wave-vector $\mathbf{k}=\left(0,0,1.5 \right)$ for $S=5/2$ Fe$^{3+}$ ions in the structural space group $R\bar{3}$ of \FB\ identifies six possible ground-states.
These states are labeled in Figs.~\ref{fig:NEUTRON}g-l by their magnetic subgroups and irreducible representations (irreps) as $R_l\bar{3}$ (irrep: $mT_{1-}$), $R_l\bar{3}$ ($mT_{1+}$), $R_l 3$ ($mT_{1-}$), $P_s\bar{1}$ ($mT_{1-}$), $P_s\bar{1}$ ($mT_{1+}$), and $P_s 1$ ($mT_{1-}$).
The first two candidates in Figs.~\ref{fig:NEUTRON}g and \ref{fig:NEUTRON}h represent maximal symmetry subgroups corresponding to out-of-plane Ising spins ($M_z\neq0$ and $M_{x,y}=0$) with A-type and C-type AFM ordering, respectively.
The lower symmetry $R_l 3$ similarly allows only $M_z$ components but with two different moment sizes (Fig.~\ref{fig:NEUTRON}i).  
The later three subgroups (Figs.~\ref{fig:NEUTRON}j,k,l) are primitive ($P$) instead of rhombohedral ($R$).
Having a lower symmetry, they allow for each of the previous cases in Figs.~\ref{fig:NEUTRON}g,h,i to have moments canted toward the $ab$-plane, i.e. $M_{x,y}\neq 0$.

We identify the ground-state of \FB\ as A-type AFM order (Fig.~\ref{fig:NEUTRON}g) for the following reasons. 
(i) The lack of magnetic reflections along the $00L$ direction in Fig.~\ref{fig:NEUTRON}e indicates that the allowed magnetic moments lie along the $c$-axis, which disqualifies the configurations in Figs.~\ref{fig:NEUTRON}j,k,l with allowed moments in the $ab$-plane.
(ii) Our neutron scattering refinement of the magnetic moment size (SI) gives 4.90(99)~$\mu_B$ on each Fe$^{3+}$ site, consistent with the results of magnetization measurements and slightly larger than 4.3~$\mu_B$ reported for \FC\ in prior studies~\cite{cable_neutron-diffraction_1962,jones_lowtemperature_1969}.
Thus, the configuration in Fig.~\ref{fig:NEUTRON}i with different moment sizes on different Fe-sites is also disqualified.
(iii) The remaining candidates in Figs.~\ref{fig:NEUTRON}g and \ref{fig:NEUTRON}h belong to the maximal subgroup $R_l\bar{3}$ corresponding to the A-type and C-type AFM ordering, respectively.
These two structures can be distinguished according to the intensity of magnetic diffraction peaks.
We simulated the structure factor ($F^2_{sim}$) for several nuclear and magnetic Bragg peaks in both A-type and C-type structures assuming $\mu=4.9~\mu_B$, and summarized them in the Supplementary Table~I (SI).
According to Table~I, $F^2_{sim}$ values for two magnetic Bragg peaks are vanishingly small in the C-type AFM structure and considerably large in the A-type AFM structure.
Thus, the diffraction pattern is best described by A-type AFM ordering (Fig.~\ref{fig:NEUTRON}g).

An A-type AFM ground-state is also consistent with the field-scale of the MM transition in Fig.~\ref{fig:MAG}c which is about 3~T in \FB\ corresponding to 2~K, considerably smaller than $T_\text{N} = 14$~K.
Thus, the MM transitions are likely due to a spin flip between the layers instead of within the layers, so the ground-state must be A-type instead of C-type AFM.

\subsection{\label{subsec:DFT}First-Principles Calculations.} 
\begin{figure}
  \includegraphics[width=0.46\textwidth]{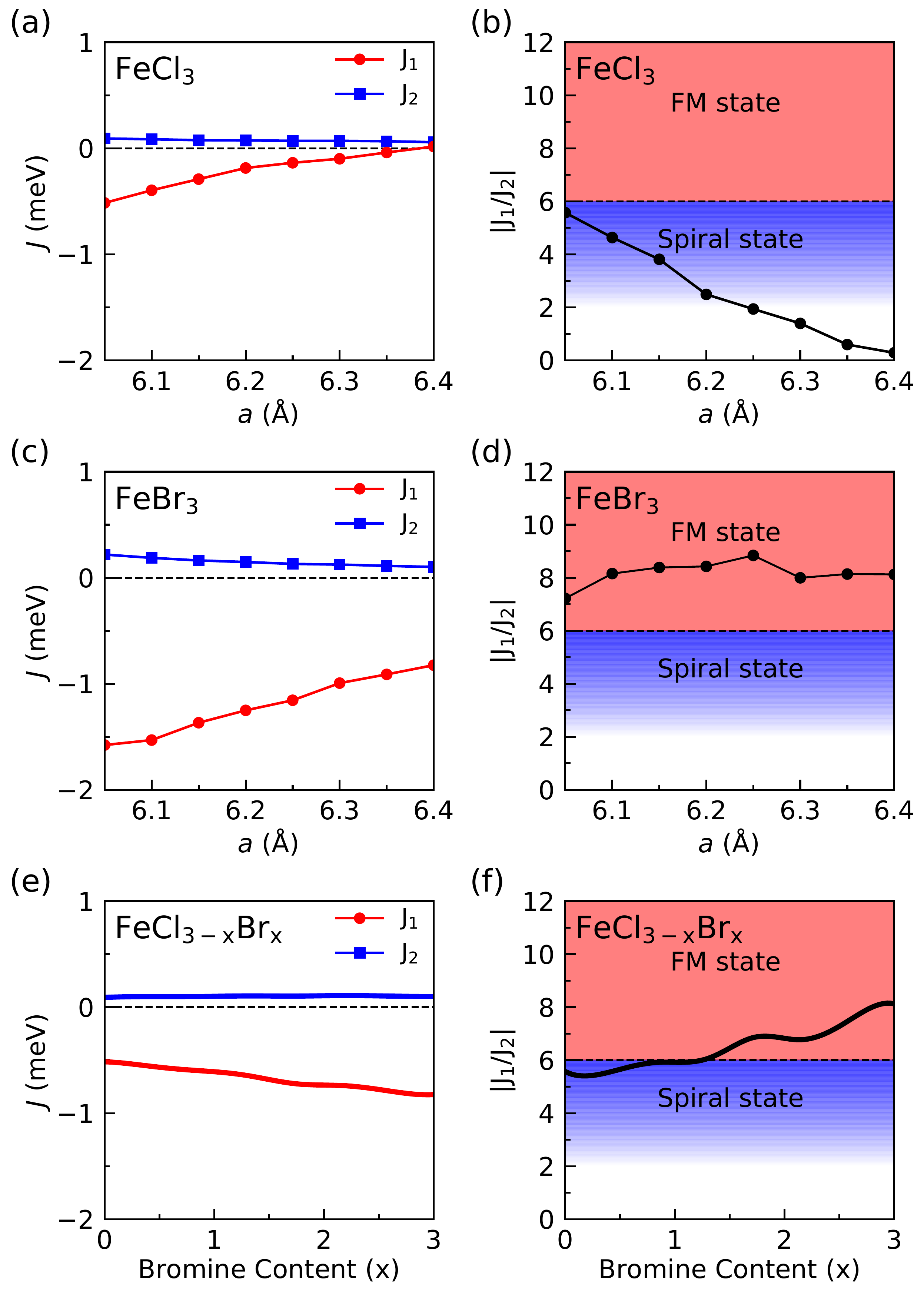}
  \caption{\label{fig:DFT}
  (a) The NN ($J_1$) and NNN ($J_2$) coupling constants computed from first-principles and plotted as a function of in-plane lattice parameter $a$ in stoichiometric \FC.
  (b) The ratio $|J_1/J_2|$ plotted as a function of $a$. 
  (c,d) Same as in (a,b) but for the stoichimetric \FB.
  (e) $J_1$ and $J_2$ traced as a function of bromine content $x$ in \FCB\ using virtual lattice approximation.
  (f) $|J_1/J_2|$ as a function of $x$.
  }
\end{figure}
Our experimental results suggest that \FC\ must be very close to a QPT since the magnetic ground-state changes from the spiral order in \FC\ to an A-type AFM order upon $3\%$ bromine doping ($x=0.08$ in \FCB). 
To understand the nature of the transition, we performed all-electron first-principles calculations based on DFT in \FC\ and \FB\ to extract $J_1$ and $J_2$ exchange couplings in the model Hamiltonian Eq.~\ref{eq:J1J2} for a 2D honeycomb lattice of $S=5/2$ spins (Fig.~\ref{fig:CIF}a).
We extracted $J_1$ and $J_2$ as a function of the in-plane lattice parameters interpolating between the experimental values of \FC\ and \FB, $a =6.05$ and $6.40$~\AA, respectively. 
As a benchmark to our methodology, we have obtained $J_1$ and $J_2$ values in good agreement with those reported for FeCl$_3$ via inelastic neutron scattering~\cite{gao_spiral_2022} giving rise to a spiral spin liquid state with $|J_1/J_2|\approx 4$.

To reproduce the high-spin configuration and $J_1/J_2$ found experimentally in \FC\ and \FB, onsite Coulomb interactions must be included in DFT.
We found that $U_{dd}=7.0$, $U_{pp}= 0.5$, and $U_{pp}= 3.5$ eV were required in the Fe, Cl, and Br atoms, respectively, to obtain the SSL state in \FC\ and in-plane FM state in \FB\ (Supplementary Figs.~S2,3).
We verified that this trend was robust against changes in the local interactions, including spin-orbit coupling effects (SI).
The larger value of $p$-orbital potential in \FB\ was necessary for establishing the experimentally observed in-plane FM state.
Smaller values of $U_{pp}$ would lead to a much larger $J_2$ and establish an AFM (N\'{e}el) state within the honeycomb planes of \FB\ due to the more covalent nature of Fe-Br bonds compared to Fe-Cl bonds.
Note that the in-plane correlations are FM within an A-type AFM order (Fig.~\ref{fig:NEUTRON}g) consistent with the positive \TW\ observed in \FB\ (Fig.~\ref{fig:MAG}f). 
Our calculations are performed on a 2D lattice without considering the inter-layer coupling $J_c$ that eventually establishes the 3D A-type AFM order in \FB\ (SI).

Figures~\ref{fig:DFT}a-d show the evolution of $J_1$, $J_2$ and $|J_1/J_2|$ as a function of the lattice parameter for \FC\ and \FB. 
We observe that for stoichiometric \FC, an increase in the lattice parameter dramatically impacts $|J_1|$ unlike $|J_2|$ (Fig.~\ref{fig:DFT}a), while the system remains in the spiral state (Fig. \ref{fig:DFT}b). 
In contrast, the stoichiometric \FB\ remains in the FM state for all lattice parameter values due to a much larger $|J_1|$ (Figs.~\ref{fig:DFT}c,d). 
The dashed lines in Figs.~\ref{fig:DFT}b,d mark the critical value $|J_1/J_2|=2Z=6$ for the theoretical transition from the SSL to FM/N\'{e}el state~\cite{bergman_order-by-disorder_2007,gao_spiral_2017,niggemann_classical_2019}.
To account for the alloys \FCB, we use a virtual crystal approximation (SI) to estimate $J_1$ and $J_2$ as a function of $x$ based on the alloy-dependent lattice constant (Figs.~\ref{fig:DFT}e,f). 
Within this approximation, doping \FC\ with Br drives the 2D system from the spiral to FM phase, as observed experimentally.  

The theoretical phase diagram in Fig.~\ref{fig:DFT}f shows a QPT between the spiral and FM states at $x=0.8$ in 2D layers of \FCB.
This is consistent with the experimental data in Figs.~\ref{fig:MAG}e,f,h that show a jump in \TN, sign change in \TW, and MM transition at $x>0.08$.
However, the theoretically predicted critical doping $x_c=0.8$ differs from the experimentally observed $x_c=0.08$.
Such a difference likely stems from limitations of the DFT methodology that neglects the impact of disorder on exchange interactions.
Spin liquid phases (e.g. SSL) are particularly sensitive to disorder~\cite{bergman_order-by-disorder_2007,niggemann_classical_2019,kao_vacancy-induced_2021,dantas_disorder_2022} and thus, it is conceivable that the alloying procedure would change the critical doping via disorder effects that are not captured in the pristine-limit DFT calculations. 

\section{\label{sec:conclue}Conclusion}
To summarize, we have demonstrated a QPT by tuning the halide composition in the frustrated VdW system \FCB.
Our results demonstrate the application of halide engineering in tuning the $J_1/J_2$ ratio of the underlying frustrated honeycomb spin model, in particular crossing the critical point of the spin spiral liquid at $x_c=0.08$. 
The transition driven by the halide composition from a spiral to FM state was demonstrated with magnetometry measurements, and the ground-states were determined by neutron scattering results presented here for \FB\ and elsewhere~\cite{cable_neutron-diffraction_1962} for \FC. 
Our theoretical calculations further show that the transition from the spiral to FM state is driven by competing magnetic exchanges
with a sizable contribution from the $p$-orbital correlations of halides. 
Our results establish a new strategy for engineering frustrated VdW magnetic materials by exploiting a continuous parameter realized by mixed halide chemistry.

\section*{ACKNOWLEDGMENTS}
The work at Boston College was supported by the National Science Foundation under the grant number DMR-2203512.
J.L.L. and A.F. acknowledge the computational resources provided by the Aalto Science-IT project, and the financial support from the Academy of Finland Projects No. 331342, No. 336243 and No 349696, and the Jane and Aatos Erkko Foundation.
A portion of this research used resources at the High Flux Isotope Reactor, a DOE Office of Science User Facility operated by the Oak Ridge National Laboratory.




\bibliography{Cole_3Mar2023}

\end{document}